\documentclass[reprint,twocolumn]{revtex4}
\usepackage[pdftex]{graphicx}
\usepackage{amsmath}
\usepackage[linkcolor=blue,citecolor=blue]{hyperref} 
\hypersetup{colorlinks=true}
\newcommand\ee{\end{equation}}
\newcommand\be{\begin{equation}}

\begin{document}

\title{Detecting CNO solar neutrinos in next-generation xenon dark matter experiments}
\author{Jayden~L.~Newstead$^{\bf a}$}
\author{Louis~E.~Strigari$^{\bf b}$}
\author{Rafael~F.~Lang$^{\bf c}$}
\affiliation{$^{\bf a}$Department of Physics, Arizona State University, Tempe, AZ 85287, USA}
\affiliation{$^{\bf b}$Mitchell Institute for Fundamental Physics and Astronomy,  Department of Physics and Astronomy, Texas A\&M University, College Station, TX 77845, USA}
\affiliation{$^{\bf c}$Department of Physics and Astronomy, Purdue University, West Lafayette, IN 47907, USA}

\begin{abstract}
We study the prospects for measuring the low-energy components of the solar neutrino flux in future direct dark matter detection experiments. We show that for a depletion of $^{136}$Xe by a factor of 1000 relative to its natural abundance, and an extension to electron recoil energies of $\sim$ MeV, future xenon experiments with exposure $\sim 1000$ ton-yr can detect the CNO component of the solar neutrino flux at $\sim 3 \sigma$ significance. A CNO detection will provide important insight into metallicity of the solar interior. Precise measurement of low-energy solar neutrinos, including as $pp$, $^7$Be, and $pep$ components, will further improve constraints on the ``neutrino luminosity'' of the Sun, thereby providing constraints on alternative sources of energy production. We find that a measurement of $L_{\nu}/L_{\odot}$ of order one percent is possible with the above exposure, improving on current bounds from a global analysis of solar neutrino data by a factor of about seven.
\end{abstract}

\maketitle

\section{Introduction} 

\par Future direct dark matter detection experiments will be sensitive to $\sim$ MeV energy neutrinos from astrophysical sources. In addition to the importance of understanding these neutrinos as a background for dark matter~\cite{Monroe:2007xp,Strigari:2009bq,Billard:2013qya}, the identification of them will provide important information on the properties of the sources and in searches for physics beyond the Standard Model~\cite{Pospelov:2011ha,Harnik:2012ni,Billard:2014yka,Dutta:2017nht}. While dark matter detectors are not optimized to measure neutrino signals, measurements can be made during, and at no cost to, their primary searches.   

\par Solar neutrinos represent a particularly interesting source that will be detected through two primary channels: coherent neutrino-nucleus scattering and neutrino-electron elastic scattering. Though less well studied in the context of dark matter detectors than the coherent scattering channel, neutrino-electron elastic scattering is important~\cite{Billard:2014yka,Schumann:2015cpa,Dutta:2017nht}, as it may be the first astrophysical neutrino signal measured in dark matter detectors. 

\par Several solar neutrino experiments have measured neutrino-electron elastic scattering, typically by examining a relatively narrow range in electron recoil energy, and thereby isolating a specific component of the solar neutrino flux. For example, Super-K~\cite{Abe:2010hy}, SNO~\cite{Aharmim:2011vm}, and Borexino~\cite{Bellini:2008mr} directly measured the $^8$B component, and the first results from Borexino identified the low-energy $^7$Be~\cite{Bellini:2011rx}, $pep$~\cite{Collaboration:2011nga}, and $pp$~\cite{Bellini:2014uqa} components. More recently, Borexino has become the first experiment to perform an analysis on multiple components of the solar neutrino flux~\cite{Agostini:2017ixy}. Their multicomponent spectral fit in the energy range 0.19-2.93 MeV provides the most precise measurements of the $pp$, $pep$ and $^7$Be fluxes. They also derived an upper bound on the CNO flux which is $\sim 4$ times larger than the predicted Standard Solar Model (SSM) flux.

\par What new information can be extracted from the neutrino-electron elastic scattering channel in dark matter detectors? Though electron scattering of solar neutrinos has been studied during the past several decades, there are still some outstanding issues that the above data, and more generally all solar neutrino data, do not conclusively address (for recent reviews see Refs.~\cite{Robertson:2012ib,Antonelli:2012qu}). For example, from an astrophysical perspective there is the often-discussed solar metallicity problem. Theoretical modeling suggests a lower abundance of metals in the solar core, i.e. a low-Z SSM~\cite{Asplund:2009fu}, in comparison to the previously established high-Z SSM~\cite{Grevesse:1998bj}. Though some solar neutrino experiments favor a high-Z SSM, a global analysis of all solar neutrino fluxes remains inconclusive. An improved measurement of the $^8$B component, and a first measurement of the low energy CNO component, will help to shed light on this issue. The SNO+ detector, presently running with light water \cite{Caden:2017htb}, would be capable of resolving the problem in as little as 5 years of tellurium-free runtime~\cite{Cerdeno:2017xxl}. However, given the current planned addition of tellurium in late 2018, it is unclear when SNO+ will be able to gain the requisite exposure.

\par In this paper we discuss the prospects for performing a multicomponent spectral analysis on the solar neutrino signal in dark matter detectors via the neutrino-electron elastic scattering channel. A similar analysis was undertaken for the nuclear recoil signal in Refs.~\cite{Strigari:2016ztv,Cerdeno:2017xxl}, which requires detectors with very low recoil energy thresholds. When utilizing the neutrino-electron scattering channel low thresholds are not required, but the lower rate does require larger exposures. We note that Refs.~\cite{Franco:2015pha,Cerdeno:2017xxl} have recently performed an analysis of the solar neutrino-electron elastic scattering signal in future larger scale argon detectors, focusing mostly on the prospects for detection of the CNO flux. They find that argon detectors with exposures of 500-1000 ton-years can solve the solar metallicity problem, depending on background assumptions. In this paper we choose to focus on xenon experiments, which are now establishing world-leading limits on dark matter over a mass range above $\sim 10$ GeV~\cite{Akerib:2016vxi,Aprile:2018dbl}. We include the associated backgrounds that at present prevent an extraction of the $pp$ neutrino-electron elastic scattering signal. We also provide an estimate of the experimental background reduction that is required for xenon experiments to extract new physics from the solar neutrino flux. 

\par We highlight two key results. First, for a plausible background reduction, xenon-based experiments sensitive to electron recoils up to $\sim$ MeV energies can make the first detection of the CNO flux component. Second, the neutrino-electron elastic scattering channel provides a means to improve the measurement of the ``neutrino luminosity'' of the Sun, and to measure the fraction of the solar energy that is generated from the $pp$ and CNO chains. These measurements are important for understanding the solar interior, and the possibility of alternative energy sources within the solar interior. 

\section{Neutrino fluxes and experimental backgrounds}
\label{sec:multi}
\par In this section we discuss the relevant features of the neutrino-electron elastic scattering rate from solar neutrinos. We work within the context of a xenon dark matter detector, outlining both the flux predictions and the experimental background rates. It is straightforward to translate this analysis to other targets (or to incorporate multiple targets), such as argon, though in general the experimental backgrounds will be different for different targets. For all the results in this section we use the high-Z SSM, with flux normalization coefficients given in Ref.~\cite{Robertson:2012ib}.

\subsection{Neutrino signals} 

\begin{figure}
\centering
  \includegraphics[width=.80\linewidth]{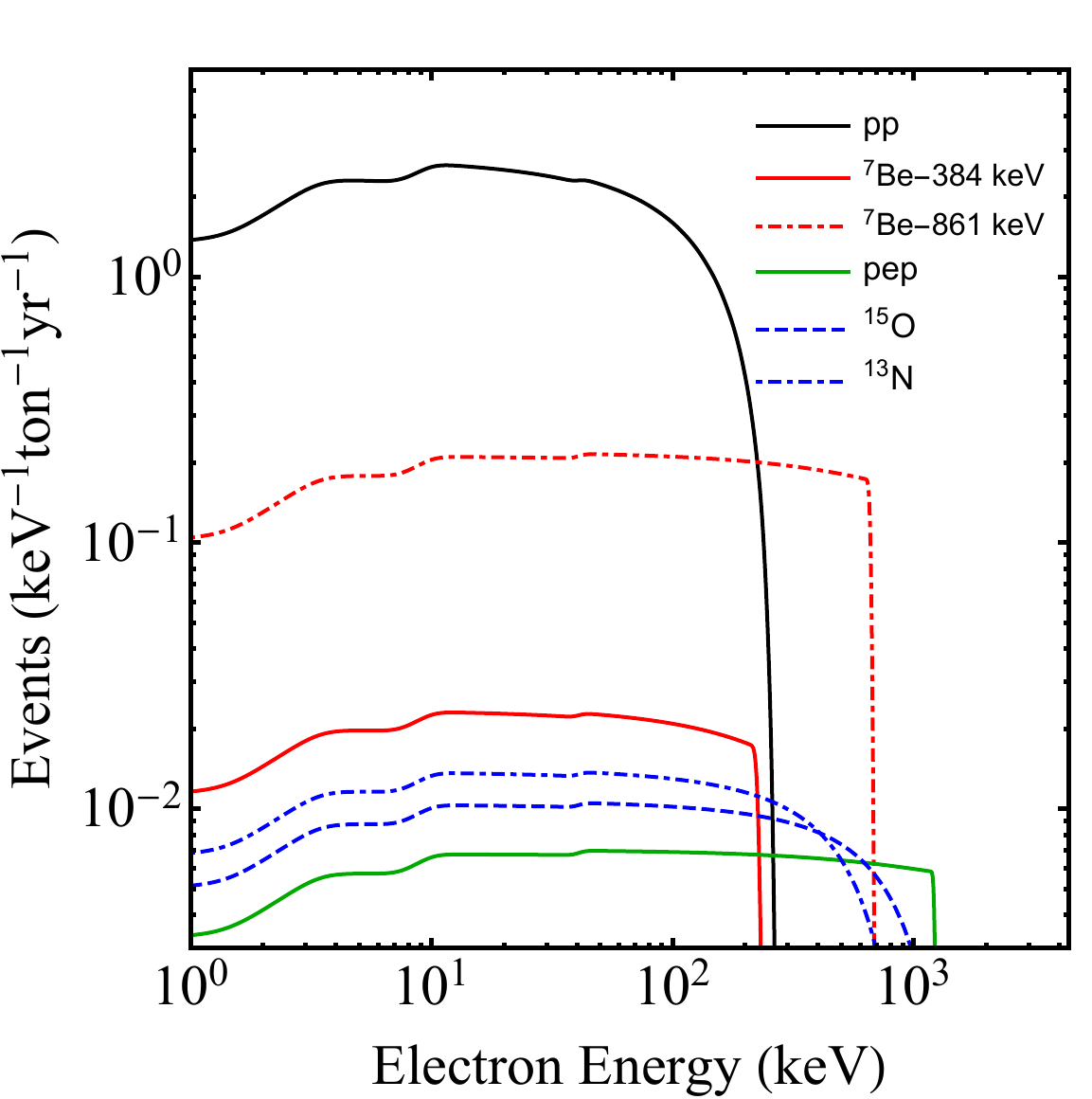} 
\caption{Electronic recoil spectrum from solar neutrinos in xenon experiments for elastic $\nu + e^-$ scattering. The labels denote the $pp$, $^7$Be, CNO, and $pep$ fluxes. The blue curves show the relevant CNO components ($^{15}$O and $^{13}$N).}
\label{fig:spectrum}
\end{figure}

\begin{figure}
\centering
  \includegraphics[width=.80\linewidth]{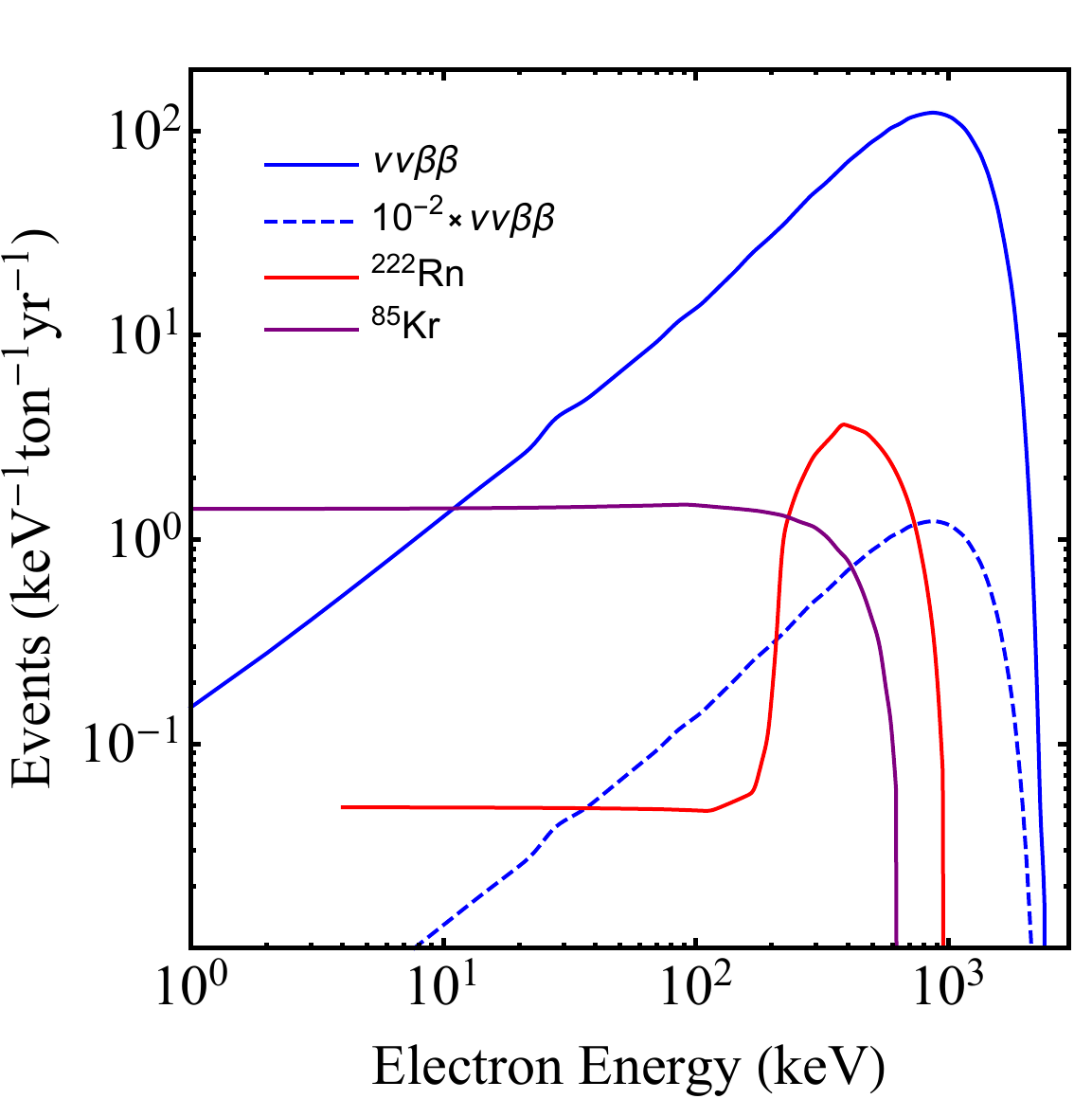} 
\caption{The background electronic recoil spectrum from sources relevant to xenon experiments. The solid blue (dashed) line is the 2$\nu \beta \beta$ decay of $^{136}$Xe, at natural abundances (depleted to $1\%$ of its natural abundance). The purple line is the background due to $^{85}$Kr at a concentration of 0.1 ppt. The red line is the spectrum from $^{216}$Pb due to $^{222}$Rn emanation, with an activity of 0.1 $\mu$Bq/kg.}
\label{fig:bgSpectrum}
\end{figure}  

\par Figure~\ref{fig:spectrum} shows the electron recoil event rate spectrum for the most prominent low-energy solar flux components: $pp$, $^7$Be (both 384 keV and 861 keV lines), $pep$, $^{15}$O and $^{13}$N. We take an electron neutrino survival probability of $P_{ee} \simeq 0.55$, which is consistent with the LMA-MSW solution in the low energy, vacuum-dominated regime~\citep{Agostini:2017ixy,Bellini:2014uqa}.  The stepping approximation is taken to account for electron energy levels, more detailed atomic effects have been neglected but may produce corrections to the rate~\cite{Chen:2016eab}. The event rates have been smoothed by a gaussian energy resolution of width (in keV) given by:
\be
\frac{\sigma(E_R)}{\mathrm{keV}} = 0.31\sqrt{\frac{E_R}{\mathrm{keV}}} + 0.0035 \frac{E_R}{\mathrm{keV}},
\ee
where $E_R$ is the recoil energy. This parameterization provides a good fit to energy resolution achieved by XENON1T.

The elastic scattering rate in xenon is dominated by $pp$ neutrinos, for which the integrated rate up to the endpoint of the spectrum is $\sim 330$ per ton per year. In the $1-10$ keV window which is the most relevant for present dark matter searches, the corresponding rate is $\sim 25$ per ton per year, where it contributes to the electronic recoil background. The $pp$ rate is theoretically well determined, and is only very weakly sensitive to the assumed solar metallicity model. 

\par After $pp$, the $^{7}$Be component of the flux is the most significant. There are two separate spectral components that contribute to the $^{7}$Be flux: a 861 keV line which has a 90\% branching fraction, and a 384 keV line which has a 10\% branching fraction. In contrast to the $pp$ flux, the $^{7}$Be flux is very sensitive to the assumed solar metallicity. Assuming the high metallicity SSM, the total $^{7}$Be neutrino flux is $5 \times 10^9$ cm$^{-2}$ s$^{-1}$. Averaging over all scattering angles (which remain unknown in a typical liquid xenon experiment), the spectra as shown in Figure~\ref{fig:spectrum} result. For the 861 keV line, the interaction rate is $\sim 129$ per ton per year integrated over all recoil energies up to the endpoint. For the 384 keV line, the corresponding interaction rate is $\sim 5$ per ton per year integrated over all energies up to the endpoint. 

\par The $pep$ and CNO spectra are the next most prominent components. The CNO spectrum is the sum of three components: $^{13}$N, $^{15}$O, and $^{17}$F. In Figure~\ref{fig:spectrum}, the plotted CNO spectrum is the sum of the $^{13}$N and $^{15}$O components, with the $^{17}$F flux contributing a negligible rate in the recoil range of interest. Note that examining Figure~\ref{fig:spectrum} the $pep$ rate is only larger than the CNO rate for recoil energies $\gtrsim 1$ MeV. Experiments such as Borexino exploit this by detecting electron recoils in a small window near the endpoint of the $pep$ spectrum. For recoil energies $\lesssim 100$ keV, the CNO spectrum is similar in shape and normalization to the $^7$Be 384~keV line spectrum. For CNO, the total rate integrated over all energies is 14 per ton per year. 

\subsection{Experimental backgrounds} 

\par Figure~\ref{fig:bgSpectrum} shows an estimate of the experimental backgrounds that are intrinsic to a xenon experiment. The most prominent background arises from the two-neutrino double beta decay (2$\nu \beta \beta$) of $^{136}$Xe. This is a rare decay process of $^{136}$Xe with a natural abundance of 8.9\%. The measured energy spectrum is the sum of the energy of the two outgoing electrons, with an endpoint of 2.459 MeV. The measured half-life is $\sim 2 \times 10^{21}$ years~\cite{Ackerman:2011gz}, which corresponds to an event rate of 5.8 events per ton per year in the dark matter search range $2-10$ keV, and $\sim 10^5$ events below 1.2 MeV. Though dominant over a large electron recoil energy range, the 2$\nu \beta \beta$ background can be reduced through the use of xenon depleted of $^{136}$Xe. Experiments such as EXO~\cite{Albert:2017owj}, which aim to observe neutrinoless double beta decay, enrich the $^{136}$Xe of their target mass. A reduction of the $^{136}$Xe concentration by a factor of 100 is readily achieved in the depleted off-gas from this enrichment process. Fortuitously, the initial quantity of natural xenon required to reach nEXO's planned enrichment roughly matches that required for such a (depleted or not) next-generation dark matter detector.

\par Another experimental background arises from radioactive krypton, as xenon is extracted from air and is contaminated with trace amounts of $^{85}$Kr. Through cryogenic separation of the xenon inventory~\cite{Aprile:2016xhi}, XENON1T has achieved the lowest levels of krypton concentration at just 0.66 ppt~\cite{Aprile:2018dbl}. The goal for the next generation, multi-ton scale xenon target is 0.1 ppt of krypton~\cite{Aalbers:2016jon}.

Finally, the noble gas radon and in particular $^{222}$Rn with a half-life of 3.8 days is a step in the uranium and thorium decay series. It continuously emanates from detector materials and readily mixes with the xenon target. The subsequent beta-decay of $^{214}$Pb yields the energy spectrum shown in Figure~\ref{fig:bgSpectrum}. While an activity of 5 $\mu$Bq/kg has been achieved in XENON1T~\cite{Aprile:2018dbl}, for a next-generation detector a level of 0.1 $\mu$Bq/kg is assumed~\cite{Aalbers:2016jon}.

\section{Detection prospects} 
\par With the above estimates for signal and background, we now move on to discussing the prospects for detection. We briefly discuss our likelihood analysis for extracting the flux signals. 

\subsection{Measuring CNO neutrinos}
We first calculate the significance of a CNO neutrino flux detection as a function of the experimental exposure. We define our likelihood function as a product of a poisson term and a gaussian term. The poisson term is made up of 50 log-spaced recoil energy bins, while the gaussian term corresponds to the nuisance parameters, $\theta$,
\be
\mathcal{L}(f_\alpha) = \left(\prod_{\imath=1}^{20} \frac{N_\imath^{k_\imath} e^{-k_\imath}}{k_\imath !}\right)\left(\prod_{\alpha}e^{\frac{(1-f_\alpha)^2}{2\sigma_\alpha^2}}\right)
\label{eq:likelihood} 
\ee 
where $N_{\imath}$ ($k_\imath$) are the predicted (observed) number of events in the $\imath^{th}$ energy bin.  The total number of events in an energy bin from all flux components is $N_\imath^{tot}$. \par We define $f_\alpha$ as the flux normalizations for each component of the solar neutrino spectrum and the relevant backgrounds that we consider: $\alpha =$ {$pp$, ${}^7$Be, $pep$, CNO} for the signals, and $\alpha = ${Kr, Rn, $2\nu \beta \beta$} for the backgrounds. The nuisance parameters are taken to be all the non-signal compoenents, and their uncertainties, $\sigma_\alpha$, are taken to be either 1\% for the detector backgrounds, or the uncertainty on the solar flux component from \cite{Bergstrom:2016cbh}. 

We consider electron recoils over the entire energy range of 5 - 1,600 keV, where the upper limit extends slightly beyond the spectral endpoint of the CNO components, giving a background-only measurement in the highest energy bin. Having such a measurement improves the signal discrimination power when there is uncertainty in the background normalization. To calculate the detection significance of a given component, we use the profile likelihood test statistic, $q_0$, which is given by $\sqrt{q_0}$ ~\cite{Cowan:2010js}. The test statistic is calculated for a simulated representative dataset (called the `Asimov' dataset, where $f_{\mathrm{CNO}}=1$), via
\be
q_0 =
\begin{cases}
   -2 \mathrm{log}  \frac{\mathcal{L}(f_{\mathrm{CNO}}=0,\hat\theta)}{\mathcal{L}(\hat f_{\mathrm{CNO}},\hat{\hat\theta})}  & f_{\mathrm{CNO}}  \geq \hat f_{\mathrm{CNO}} \\
   0	& f_{\mathrm{CNO}} < \hat f_{\mathrm{CNO}} \\
\end{cases}
\label{eq:loglike}
\ee
where the hatted parameters denote maximization.

\par
\begin{figure}
\centering
\begin{tabular}{c}
  \includegraphics[width=.80\linewidth]{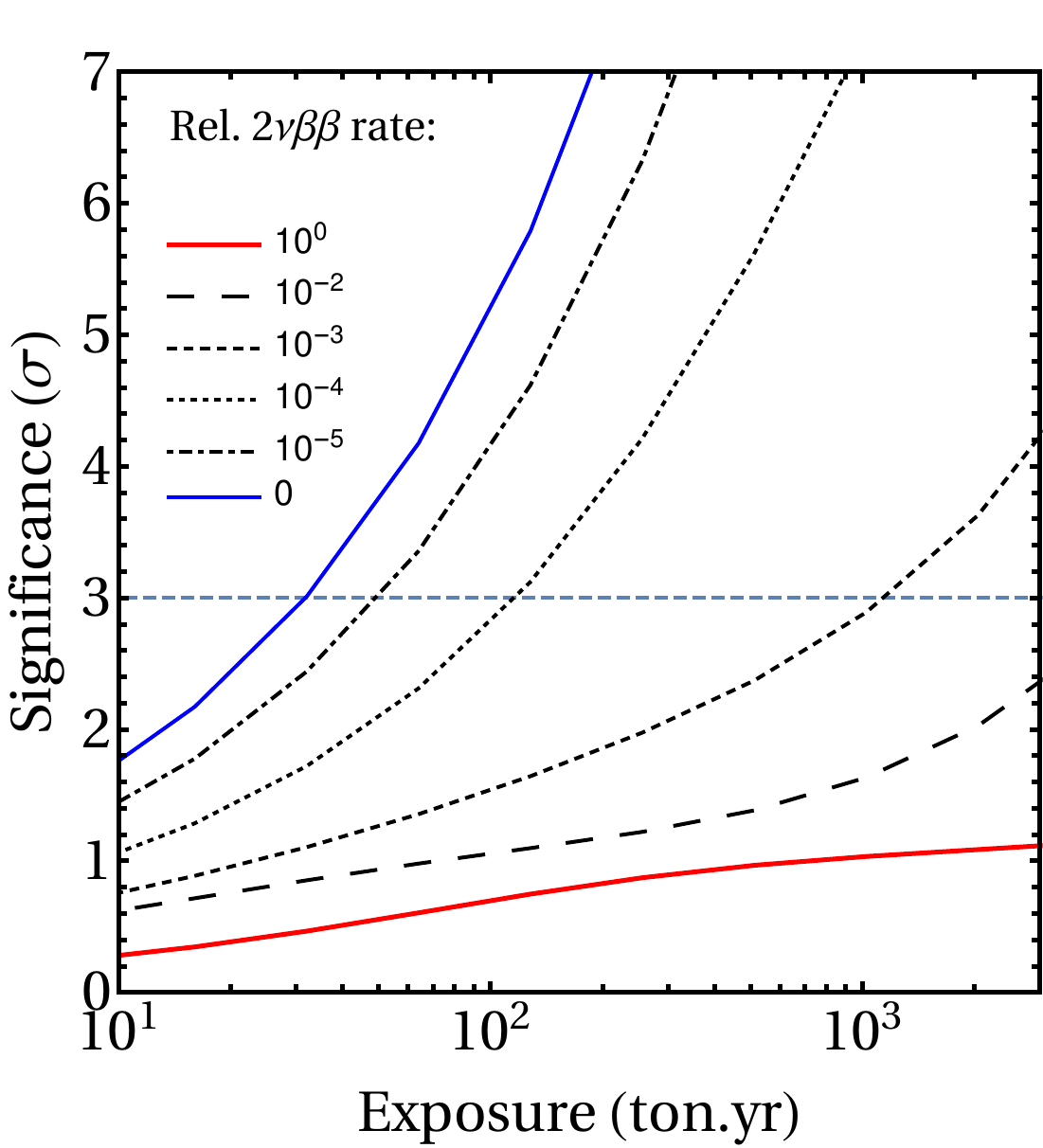} \\
  \includegraphics[width=.80\linewidth]{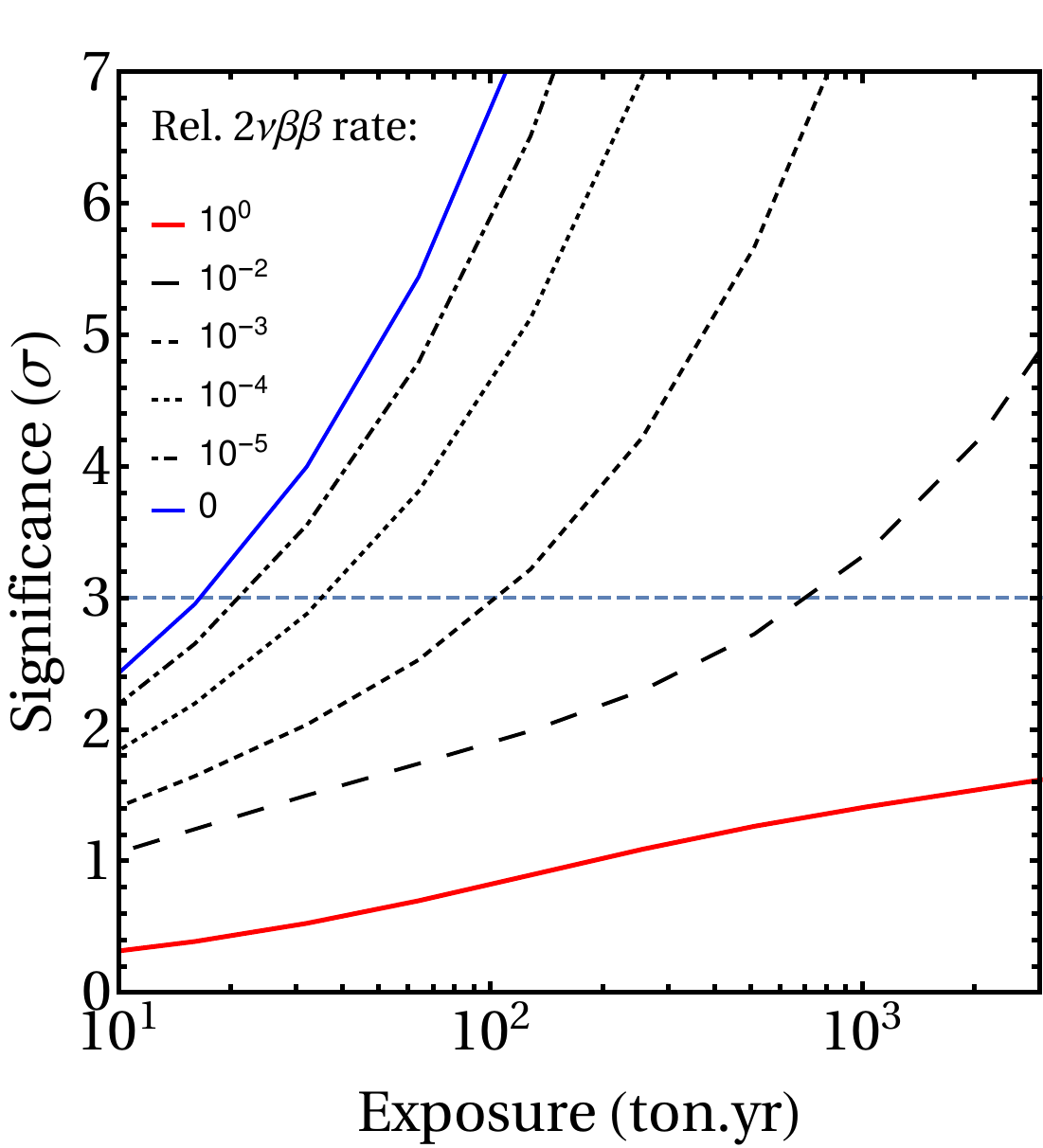} \\
\end{tabular}
\caption{The significance of a CNO neutrino detection above the background of $pp$, $pep$, $^7$Be, solar neutrinos as well as the intrinsic electron recoil background spectrum. The different lines correspond to the level of depletion of $^{136}$Xe relative to the natural abundance. The top panel assumes uncertainties of 1\% for the intrinsic backgrounds, while the bottom panel assumes 0.1\% and a factor of 10 lower concentration of Kr and Rn.
}
\label{fig:sig}
\end{figure}

 With this likelihood and test statistic, Figure~\ref{fig:sig} shows the level of significance expected for a detection of CNO neutrinos as a function of detector exposure. The significance is calculated for a series of background scenarios with progressive levels of depletion of $^{136}$Xe, and at two levels of Kr and Rn contamination. At the present projected background levels (top panel of Figure~\ref{fig:sig}), a three-sigma detection of CNO neutrinos is not possible. However, a reduction of the $^{136}$Xe concentration by a factor of $10^3$ makes such a detection possible, and further reductions can bring the exposure to feasible levels. With the simultaneous reduction in concentration of Kr and Rn by a factor of 10 (bottom panel of Figure~\ref{fig:sig}), depletion of $^{136}$Xe beyond $10^{-3}$ lowers the required exposure to levels achievable by DARWIN.

\subsection{Measuring the flux components}
\par We now move on to project uncertainties on all of the solar flux components, and for this we switch over to a Bayesian formalism. We again define a poisson likelihood function in 20 log-spaced recoil energy bins. The Bayesian priors on the five dimensional flux normalization space ($f_{pp}$, $f_{Be}$, $f_{pep}$, $f_O$ and $f_N$) are taken to be linear over a large range which encompasses the viable parameter space for each flux. In addition to the flux normalizations, we allow for the normalizations of the 2$\nu \beta \beta$, $^{85}$Kr and $^{222}$Rn to vary, and consider gaussian priors on the background flux normalizations of up to 10\%.  Prior constraints can also be applied through the consideration of the nuclear reaction chain, as applied in \cite{Bergstrom:2016cbh}; these will be explored in the following section.

\par
\begin{figure*}[ht!]
\centering
  \includegraphics[width=18cm]{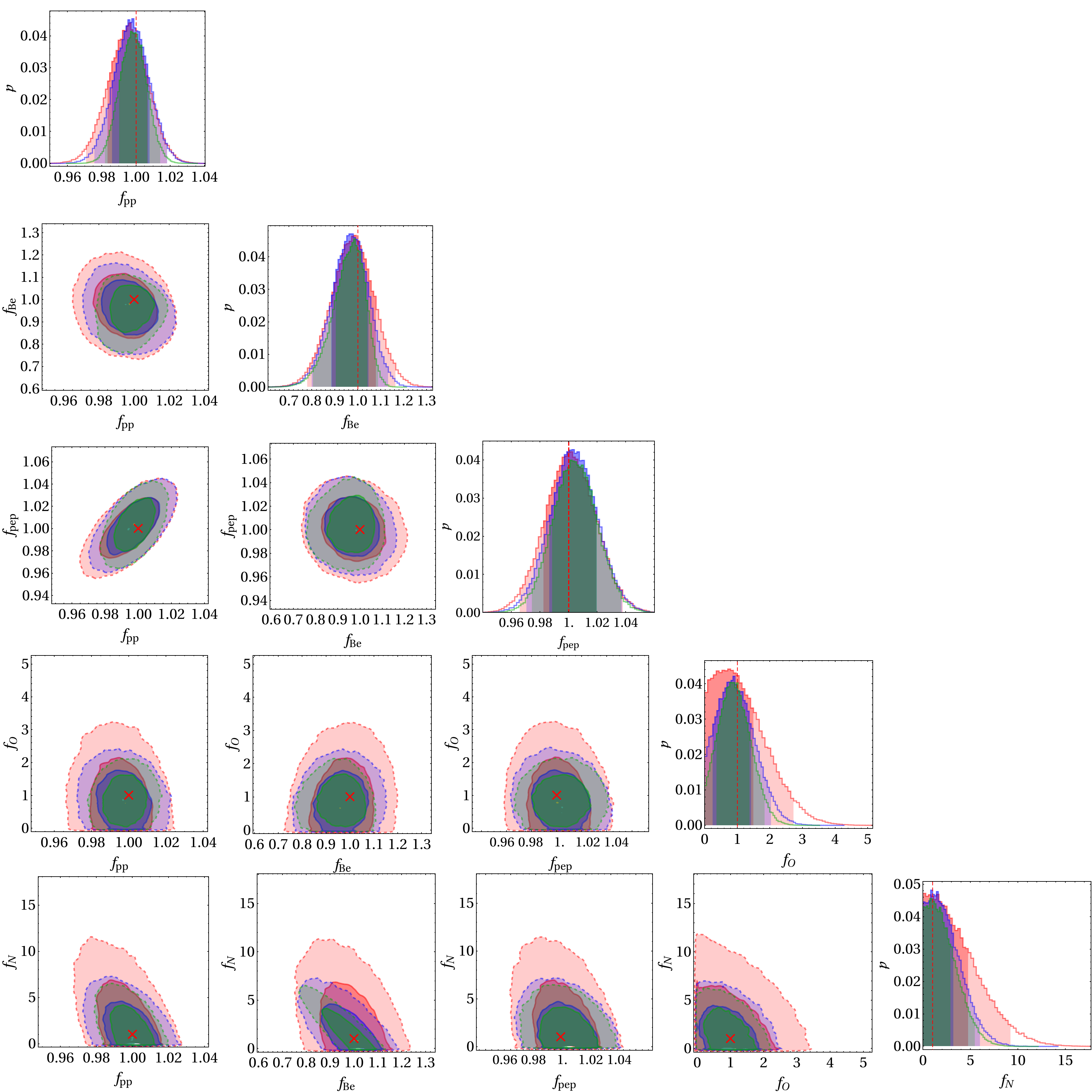} 
\caption{ Conditional posterior distributions of the flux normalizations for a 200 ton-year exposure. The different colored contours show different assumptions for the gaussian priors on the Rn, Kr, and 2$\nu \beta \beta$ background components: 10\% (red), 1\% (blue), and no background uncertainty (green). For all cases, we take the normalization of the 2$\nu \beta \beta$ background to be depleted by $10^{-3}$ relative to the natural abundance. The contours denote one and two-sigma credible regions. The red crosses and vertical lines denote the simulated values, i.e. $f_\alpha=1$.  
}
\label{fig:largeFluxPDF}
\end{figure*}

 We consider two future multi-ton scale detector scenarios: a 200 ton-year exposure is chosen to reflect a next-generation xenon detector such as DARWIN, while a 2000 ton- year exposure is chosen to represent what could ultimately be measured by a xenon detector of this type. The background spectrum of Figure~\ref{fig:bgSpectrum} is used with $^{136}$Xe depletion of 1\% to represent an achievable level of suppression, and 0.1\% to show what is achievable when the $2\nu\beta\beta$ background is subdominant. Representative datasets (where the expected number of events are realized) are generated using the central values for all flux and background normalization parameters. These events are binned into 20 log-spaced bins between 5 and 1600 keV. Fewer bins were used to save computation as it was found that increasing bin number did not improve the results significantly. These datasets are then used as the observed events in Eq.~\ref{eq:loglike}, and Bayes' theorem is used to derive constraints on the flux parameters. To sample the posterior distribution we make use of MultiNest~\cite{Feroz:2008xx}, which implements a robust multi-modal nested sampling algorithm. We marginalize over over all parameters that are not shown in a given plot in Figure's~\ref{fig:largeFluxPDF} and \ref{fig:lum}. 

\par Figure~\ref{fig:largeFluxPDF} shows the conditional probability density functions (PDFs) for an exposure of 200 ton-year. 
We consider cases in which the priors on the background components are 10\% and 1\%. We also consider the case in which there is no uncertainty on the background components in order to demonstrate the achievable reach given sufficient experimental control over those backgrounds. The uncertainty on the $pp$ and $pep$ fluxes is expected to be at the few-percent level, while the uncertainty on the $^7$Be flux is predicted to be $\sim 10\%$. Figure~\ref{fig:largeFluxPDF} also constrains the CNO fluxes to be within factor of a few, with a strong correlation between them. This correlation means that we can make a better measurement of the combined CNO flux than we can of the individual fluxes, which we will see in the following section. Note that there is a degeneracy in determination of these fluxes because the electronic recoil energy distribution is similar for both of these fluxes, with the {\it pep} ($^7$Be) spectrum extending to a slightly higher (lower) endpoint than the CNO. 

\subsection{Luminosity constraints}
\par Since nuclear fusion is the dominant source of energy that powers the Sun, a linear combination of the neutrino fluxes must add to the flux expected from the total photon luminosity. If the linear combination of neutrino fluxes, i.e. the ``neutrino luminosity'' of the Sun, deviates from the measured solar luminosity, this may be a hint of non-standard sources of energy generation in the Sun. The neutrino luminosity of the Sun is obtained from neutrino fluxes using~\cite{Bahcall:2001pf}
\begin{equation}
L_\nu = \sum_\imath \alpha_\imath \phi_\imath
\label{eq:Lnu}
\end{equation}
where the $\alpha_\imath$'s are coefficients that determine the contribution of the neutrino flux components to solar energy production. Equation~\ref{eq:Lnu} may be further divided into contributions from the {\it pp} and CNO cycle, $L_\nu = L_{\nu,pp} + L_{\nu,\mathrm{CNO}}$, providing a measurement of the fraction of the energy produced in the {\it pp} and CNO cycles. 

\par The best limit on the neutrino luminosity of the Sun comes from a global analysis of all solar neutrino data, which indicates that the neutrino luminosity is consistent with the Solar luminosity, $L_\nu / L_\odot = 1.04_{-0.07}^{+0.08}$~\cite{Bergstrom:2016cbh}. The corresponding fractions of the luminosities from the $pp$ and CNO chains are, respectively, $L_{\nu,pp} / L_\odot = 1.03_{-0.07}^{+0.08}$ and $L_{\nu,\mathrm{CNO}} / L_\odot = 0.008_{-0.004}^{+0.005}$. These results do not include any prior relation between any of the fluxes. Adding priors from the nuclear reaction chains such as the relation between the $pp$ and $pep$ fluxes reduces the uncertainties by $\sim 50\%$. At one-sigma the central values remain consistent between the different analyses.

\par Since the neutrino luminosity is derived directly from the neutrino fluxes, we can use the formalism in the previous section to translate projected constraints on the neutrino flux measurements to projected constraints on the neutrino luminosity, $L_{\nu}$, and the contribution from CNO, $L_{\nu,\mathrm{CNO}}$. Given that our simulated dataset used values for the flux components from the high-Z SSM model, the values inferred should be centered around $L_\nu / L_\odot = 1.001$ and $L_{\nu,\mathrm{CNO}} / L_\odot = 0.00685$.

\par
\begin{figure*}
\centering
  \begin{tabular}{cc}
    \includegraphics[width=.4\linewidth]{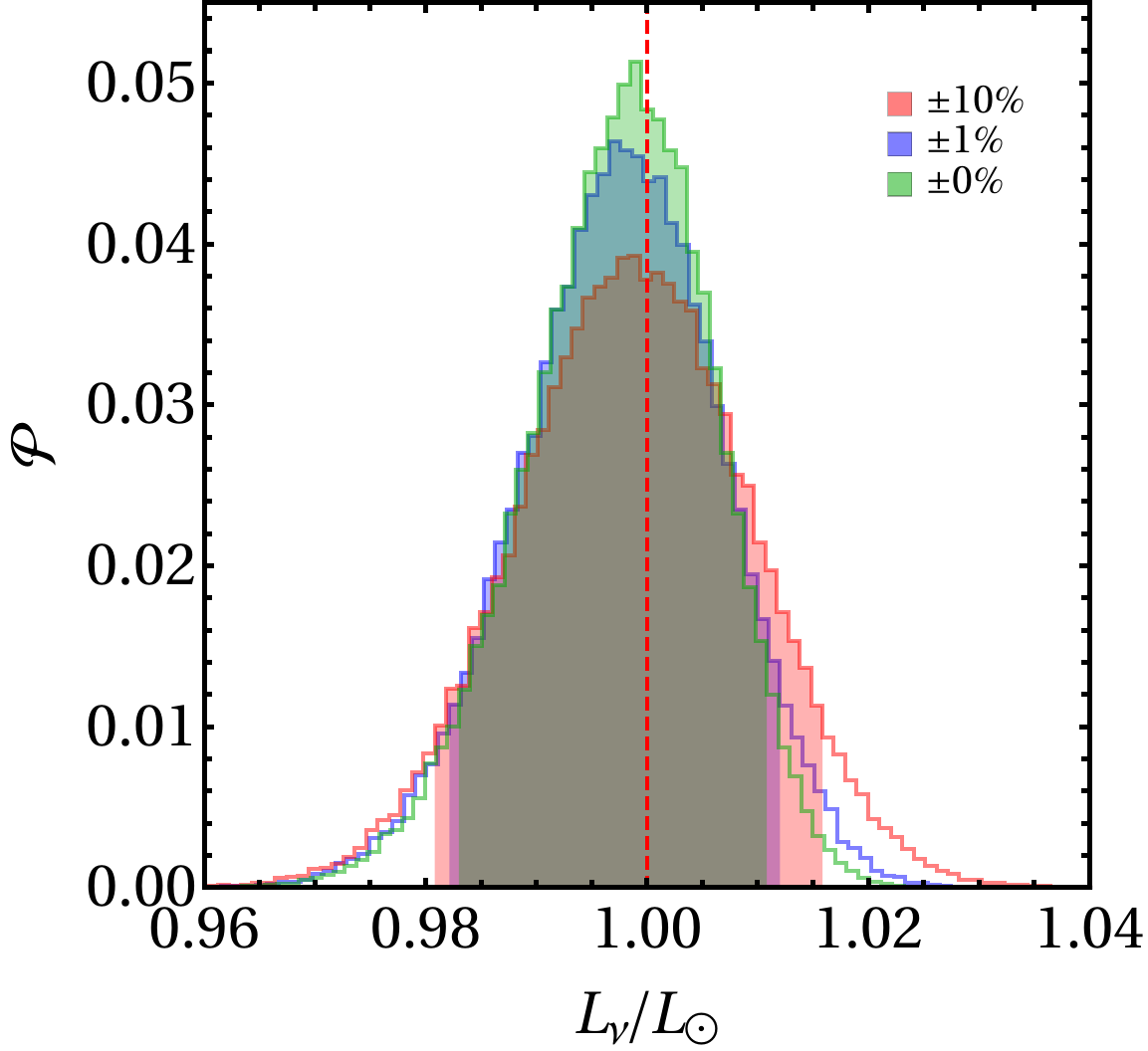} & 
    \includegraphics[width=.4\linewidth]{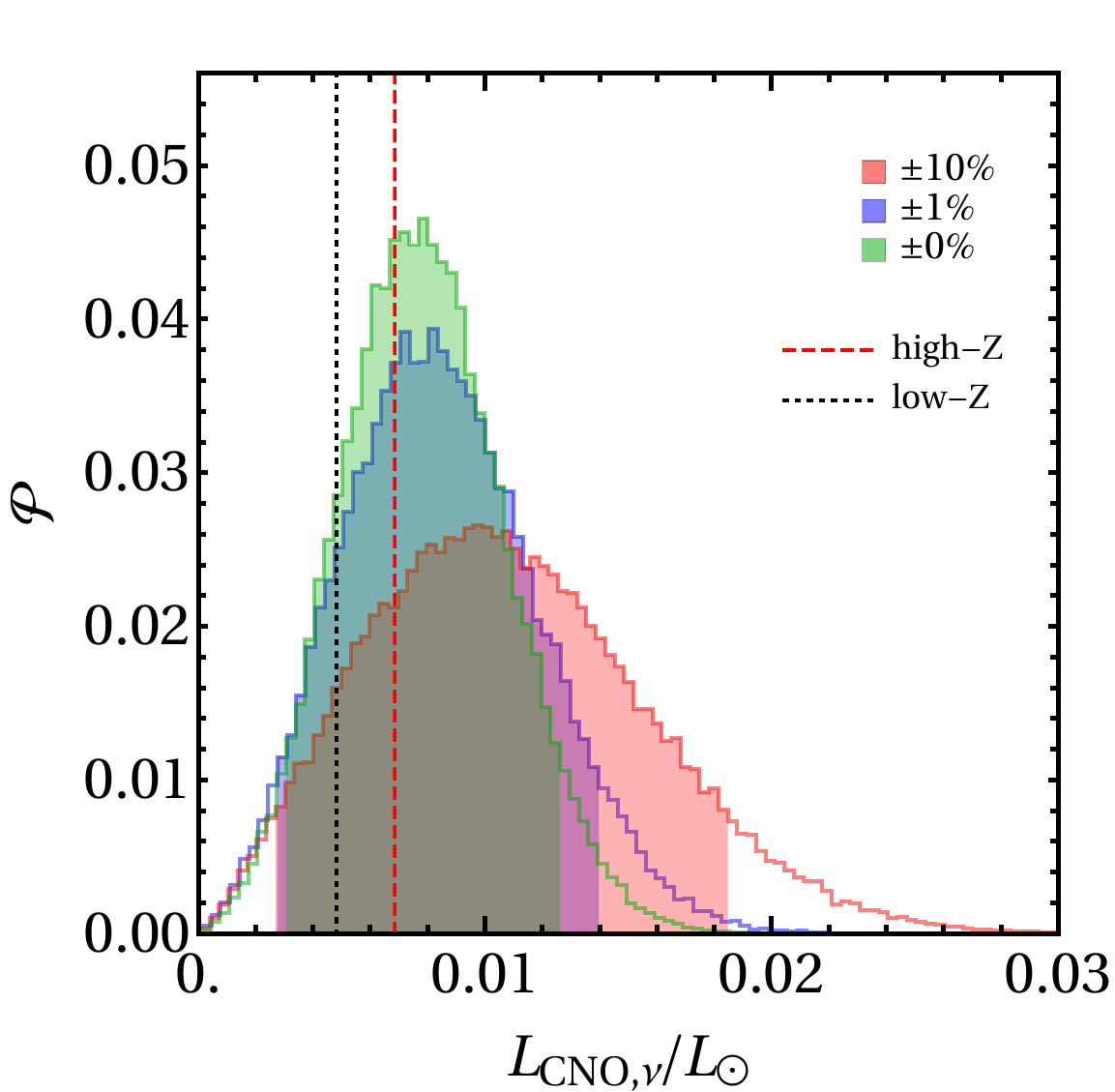} \\
    \includegraphics[width=.4\linewidth]{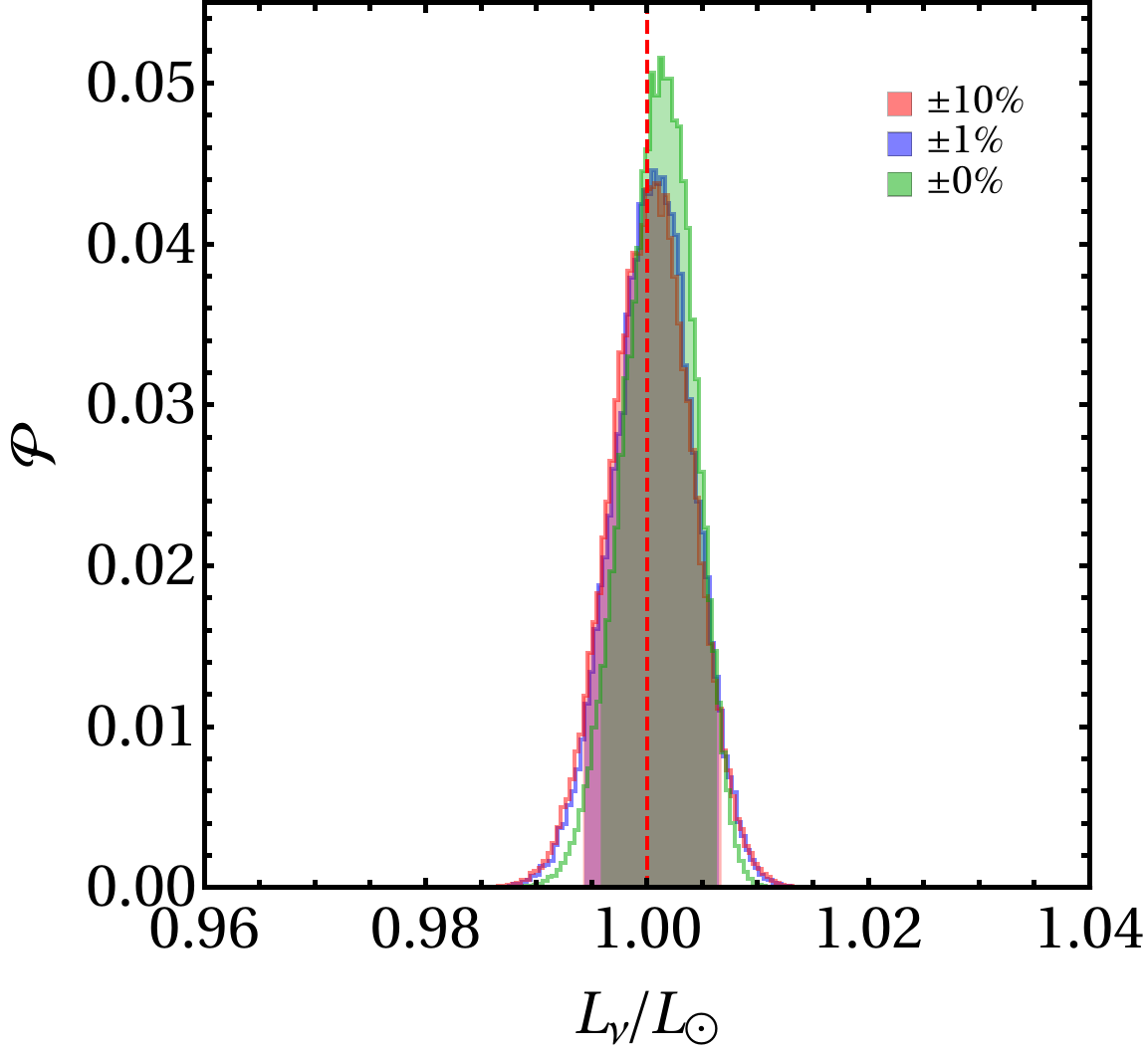} & 
    \includegraphics[width=.4\linewidth]{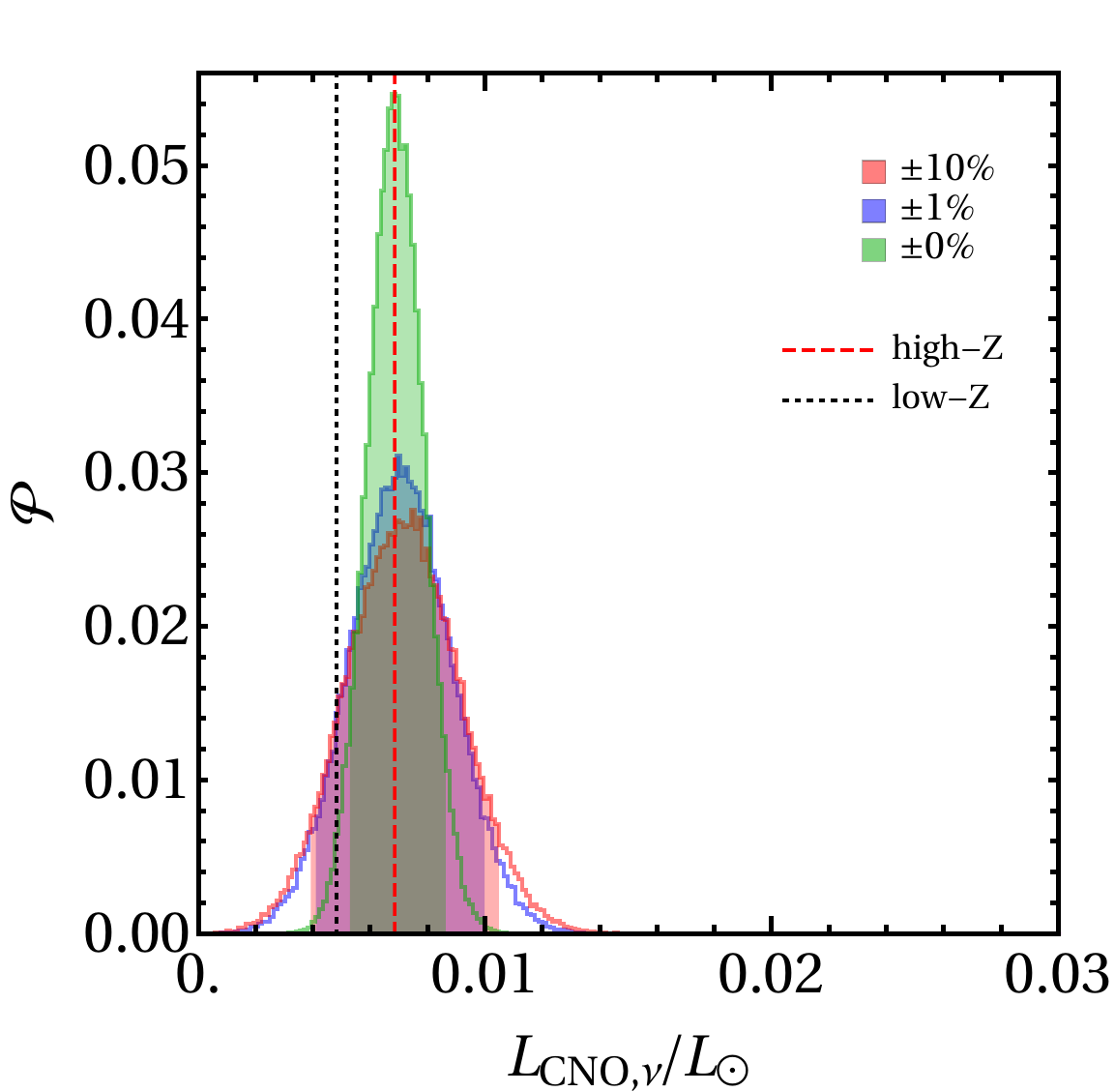} \\
  \end{tabular}
\caption{Projected fractional uncertainty on the neutrino luminosity of the Sun from multi-ton xenon dark matter experiments. The top row shows the projected constraints for a 200 ton-year exposure, and the bottom row shows the projected constraints for a 2000 ton-year exposure. The left column shows the projected constraints on the total neutrino flux, while the right column shows just the contribution of the CNO flux to the solar luminosity. The shaded regions denote the 90$\%$ credible regions. For all cases, the normalization of the 2$\nu \beta \beta$ background is depleted by $10^{-3}$ relative to the natural abundance.}
\label{fig:lum}
\end{figure*}

 Figure~\ref{fig:lum} shows the marginal PDFs of the neutrino luminosity and the fractional CNO contribution. As shown in Table~\ref{tab:results}, for the exposures that we consider, the projected constraints on $L_\nu$ are at the percent level, which is about a factor of 7 stronger than the present bounds~\cite{Bergstrom:2016cbh}. Given that $L_{\nu,CNO}$ is less than $1\%$ of the total luminosity, a reasonable estimation of its size cannot be made until $L_{\nu}$ is known at the percent level. To constrain $L_{\nu,CNO}$ to within a factor of 2, we find that exposures greater than 200 ton-years are required, where the $2\nu\beta\beta$ background has been depleted to 1\% and its normalization known to within 10\%. This result can be improved by increasing exposure, lowering background uncertainties, and further depleting the $2\nu\beta\beta$ background, see Table~\ref{tab:results}. 

\par Theoretical priors on the flux components may improve the bounds obtained in Figures~\ref{fig:largeFluxPDF} and~\ref{fig:lum}. We specifically consider a theoretical prior on the {\it pep} flux, and constraints on the relative magnitudes of the other fluxes. Since the {\it pep} reaction and the {\it pp} reaction are derived from the same nuclear matrix element, the uncertainty on the ratio of these two rates is theoretically-constrained to high precision. With this motivation we take $f_{pep}/f_{pp} = 1.006 \pm 0.013$, which represents the average of this ratio for the low and high-$Z$ SSMs~\cite{Bergstrom:2016cbh}. Motivated by the fact that the {\it pep} and {\it pp} reactions feed the $^7$Be and $^8$B reactions, and that the oxygen reaction is the slowest in the CNO process, we additionally enforce the following inequalities~\cite{Bergstrom:2016cbh},
\begin{align}
8.49\!\times\!10^{-2} f_{\mathrm{Be}} + 9.95\!\times\!10^{-5} f_{B} & \leq f_{pp} + 2.36\!\times\!10^{-3} f_{pep} \nonumber\\
f_O & \leq 1.34 f_N. \nonumber\\
\end{align}
The prior is implemented by assigning a probability of zero to points which do not satisfy these inequalities, and a gaussian factor for those that do:
\be
\pi(\theta) = 
\begin{cases}
\mathrm{exp}\left[-\frac{(f_{pep}/f_{pp} - 1.006)^2}{2 \times 0.013^2}\right] & (5)\,\,\mathrm{satisfied} \\
0 & \mathrm{otherwise.} \\
\end{cases}
\label{eq:prior}
\ee
The normalization of this prior is not important for our purposes.  

\par Table~\ref{tab:results} shows the projected constraints on the total neutrino flux and CNO fraction, both with and without the nuclear prior. The prior has a modest affect on the total neutrino luminosity measurement, most pronounced in the 2000 ton-year exposure where statistical uncertainties have been reduced. The inclusion of the priors has no impact on the inference of the CNO fraction of the neutrino luminosity. After 2000 ton-year the most optimistic CNO luminosity fraction obtainable is estimated to be $L_{\nu,\mathrm{CNO}}/L_{\odot} = (7.0\pm1.6) \times 10^{-3}$.

\begin{table}[ht]
\caption{Percentage error (at $90\%$ credible level) in measurements of the total neutrino luminosity and CNO fraction for all scenarios considered. Where the PDFs were asymmetric, the mean of the positive and negative errors was taken.}
\begin{tabular}{|r|r|r|r|r|r|r|r|}
\hline
 & & \multicolumn{2}{|c|}{$1\%$ $ 2\nu\beta\beta$} & \multicolumn{2}{|c|}{$0.1\%$ $ 2\nu\beta\beta$} & \multicolumn{2}{|c|}{$0.1\%$ $ 2\nu\beta\beta$} \\    
          Exp. &   $\%$ BG & \multicolumn{2}{|c|}{no prior} & \multicolumn{2}{|c|}{no prior} & \multicolumn{2}{|c|}{prior} \\\cline{3-8}
 (ton-yr) &  uncer. & $\sigma_{L_{CNO}}$  & $\sigma_{L_{\nu}}$ & $\sigma_{L_{CNO}}$  &  $\sigma_{L_{\nu}}$ & $\sigma_{L_{CNO}}$  & $\sigma_{L_{\nu}}$   \\
\hline
200   & $10\%$ & $89\%$ & $2.0\%$ & $75\%$ & $1.7\%$ & $75\%$ & $1.7\%$ \\
\hline
200   &  $1\%$ & $92\%$ & $1.7\%$ & $66\%$ & $1.5\%$ & $66\%$ & $1.4\%$ \\
\hline
200   &  $0\%$ & $79\%$ & $1.5\%$ & $59\%$ & $1.4\%$ & $58\%$ & $1.3\%$ \\
\hline
2000  & $10\%$ & $70\%$ & $0.74\%$ & $46\%$ & $0.60\%$ & $46\%$ & $0.58\%$ \\
\hline
2000  &  $1\%$ & $70\%$ & $0.73\%$ & $41\%$ & $0.58\%$ & $42\%$ & $0.54\%$ \\
\hline
2000  &  $0\%$ & $46\%$ & $0.63\%$ & $23\%$ & $0.49\%$ & $23\%$ & $0.42\%$ \\
\hline
\end{tabular}
\label{tab:results}
\end{table}

\section{Discussion and Conclusion} 
\label{sec:disc}
\par We have studied the prospects for measuring the low-energy components of the solar neutrino flux in future xenon direct dark matter detection experiments. For a depletion of $^{136}$Xe by a factor of $10^3$ relative to its natural abundance and an extension to electron recoil energies of $\sim$ MeV, future exposures of $\sim 1000$ ton-yr can detect the CNO component of the solar neutrino flux at $\sim 3 \sigma$ significance. This detection will provide important insight into metallicity of the solar interior. We have also shown that a precise measurement of low-energy solar neutrinos will improve bounds on the neutrino luminosity of the Sun, thereby providing constraints on alternative sources of energy production. We find that a measurement of $L_{\nu}/L_{\odot}$ of order one percent is possible with the above exposure, improving on current bounds from a global analysis of solar neutrino data by a factor of about seven. While a measurement of $L_{\nu,CNO}/L_{\odot}$ to within a factor of 2 is possible with a 200 ton-yr exposure, the measurement could be greatly improved with the depletion of $^{136}$Xe by a factor of 1000 and suppression of background uncertainties to $<\!1\%$. 

\par In this paper we considered a high-Z SSM which predicts a CNO flux that is around 25-30\% larger than the low-Z model. Thus, to be able to discern these models, an equivalently low-uncertainty measurement is required. In this paper we have demonstrated that xenon experiments with exposures of 2000 ton-years could potentially reach this benchmark, under optimistic background scenarios where the $^{136}$Xe contribution is subdominant. This is in contrast with the argon analysis of Ref.~\cite{Cerdeno:2017xxl}, which found that, optimistically, less than 1000 ton-years would be required. This shorter exposure is consistant with our findings, given their assumption of a lower level of radon contamination (10$\mu$Bq per 100 ton, versus our assumed 10$^4\mu$Bq per 100 ton). These large exposures imply 10+ year runtimes, which implies more time for other experiments to potentially make the first measurement of the CNO flux. Even so, with both xenon and argon detectors capable of making the same measurement, future dark matter experiments may provide important confirmation of such a result.

\par Measurement of neutrinos produced from the CNO cycle is a long-standing goal of the solar neutrino program. Though the CNO cycle accounts for only a small fraction of the total solar energy production, the abundance of CNO elements has important implications for the radiative transfer calculations and spectral line formation. The most recent Borexino results imply an upper bound on the CNO neutrino flux that is about a factor of four times larger than is predicted in the high-Z SSM, so is unable to test models of the solar interior. It is important to note that more well-measured components of the solar neutrino flux that are sensitive to the metallicity of the solar interior, such as $^8$B and $^7$Be, are also unable to discriminate between the low and high-Z SSMs. For this reason a measurement of the CNO flux in future solar neutrino detectors is critically important for better understanding the interior structure of the Sun. 

\par By improving constraints on the neutrino luminosity of the Sun, future detectors will also be sensitive to energy produced from hypothetical new particles. For example Weakly-Interacting Massive Particles may be a source of energy production via their accumulation and annihilation in the center of the Sun. Light particles, such as axions or hidden photons, may be thermally produced and carry away energy from the solar interior. A global analysis combining helioseismology and solar neutrino data currently place important upper limits on the couplings of these light particles~\cite{Vinyoles:2015aba}.  

\par The results in this paper highlight the importance of the development of multi-purpose detectors that are sensitive to both dark matter and astrophysical neutrinos. In the nuclear recoil channel, xenon and argon-based detectors at exposures we have considered will be sensitive to atmospheric and supernova neutrinos, with the potential to provide important new astrophysical information on these sources~\cite{Lang:2016zhv}. The depletion of $^{136}$Xe in future detectors as we have discussed makes important connections to future searches for neutrinoless double-beta decay~\cite{Albert:2017owj}. 

\bigskip 

{\bf Acknowledgements}:
The work of LES is supported by DOE Grant de-sc0010813 and that of RFL by NSF grant PHY-1719271.

\end{document}